\documentclass[12pt]{article}

\usepackage{newtxtext,newtxmath}
\usepackage{graphicx}
\usepackage{lineno}

\usepackage[letterpaper,margin=1in]{geometry}

\renewenvironment{abstract}
	{\quotation}
	{\endquotation}

\date{}

\makeatletter
\renewcommand{\fnum@figure}{\textbf{Figure \thefigure}}
\renewcommand{\fnum@table}{\textbf{Table \thetable}}
\makeatother

\usepackage{scicite}

\usepackage{url}

\def\scititle{Tempests in the Troposphere: Mapping the Impact of Giant Storms on Jupiter’s Deep Atmosphere}
\title{\bfseries \boldmath \scititle}

\author{Chris Moeckel,$^{1\ast}$ Huazhi Ge,$^{2}$ Imke de Pater$^{1,3}$\\
\small$^{1}$Department of Earth and Planetary Science, University of California, Berkeley, 94720, USA.\and 
\small$^{2}$Division of Geological and Planetary Sciences, California Institute of Technology, Pasadena, 91125, USA.\and 
\small$^{3}$Department of Astronomy, University of California, Berkeley, 94720, USA. \\  
\small$^\ast$To whom correspondence should be addressed; E-mail:  chris.moeckel@berkeley.edu 
}

\begin{document} 

% Insert the title and author list
\maketitle

% Abstract, in bold
% There are strict length limits, and not all formats have abstracts.
% Consult the journal instructions to authors for details.
% Do not cite any references in the abstract.
\begin{abstract} \bfseries \boldmath
Storms are emerging as key drivers in shaping hydrogen-dominated atmospheres.  Trace gas condensation can suppress convection and disrupt the distribution of energy and material in hydrogen atmospheres. On Jupiter, the presence of water has been invoked to control the occurrence of large-scale storms; however, the impact of storms on the ammonia and temperature distribution is unknown. We use Juno Microwave Radiometer observations of a large-scale storm in 2017 to study the aftermath of such a storm on the atmosphere. Anomalies in the retrieved ammonia abundance and atmospheric temperature show how storms deplete and heat the upper atmosphere while simultaneously depositing material well below the layers they were triggered at. These observations, aided by simulations, show that the water and ammonia cycles are coupled and that their combined effect plays a key role in explaining the depletion of ammonia in the tropospheres of Jupiter and Saturn. 
\end{abstract}

\textbf{Teaser}
Jupiter’s storms dive deep: Juno reveals how giant storms dry the upper atmosphere and deposit material in the deep atmosphere.\newline
\vspace{1cm}
% The first paragraph of any Science paper does NOT have a heading
% Nor is it indented

\section*{Introduction}\label{sec:I}
\noindent Our understanding of the redistribution of material and energy in Jupiter's troposphere and its effect on the planet's global atmospheric dynamics remain limited. Incoming solar insulation and internal heat flux contribute comparably to the atmosphere’s energy balance \cite{Hanel1981,Pirraglia1984}, resulting in a system that is forced from the top and the bottom. In Jupiter's troposphere, the incoming solar radiation is scattered and absorbed at the top of the atmosphere; however the shallow nature of the forcing ( P$<$6 bar) is in contrast with the atmospheric dynamics and variability at the pressures at which large-scale storms erupt (P $\approx$ 6 bar) and below.
% Talk about VLA 
Radio observations in recent decades have revealed a two-layer system in the troposphere of Jupiter, where trace gases do not follow the behavior predicted by thermochemical equilibrium models \cite{li2017,dePater2019b,moeckel2023}. In particular, the global depletion of ammonia (cloud condensation pressure at 0.7 bar) down to 30 bar cannot be explained with our current understanding of chemistry or Jovian atmospheric dynamics. Storms have emerged as a potential explanation \cite{Showman2005,Li2015,Guillot2020,Li2023}; however, little is known about how deep these storms can disturb the atmosphere. Since radio observations are sensitive to the combination of trace gas and temperature anomalies, they represent our best window into the dynamics of the atmosphere. In this work, we retrieve variations in the ammonia abundance or physical temperature anomalies and trace the effect of the storms below the visible cloud deck for the first time. 

% Storm outbreak details 
In December 2016, amateur astronomer Phil Miles captured the onset of a storm system in the South Equatorial Belt (SEB) (14.5$^\circ$ deg S, planetocentric), $\sim$60$^\circ$ East of the Great Red Spot.  The timing coincided fortuitously with a ground-based, multi-wavelength observation campaign to support the Juno perijoves (PJ) encounters. Observations with the Atacama Large (sub)-Millimeter Array showed that the base of the water cloud is the likely region from which the storms originated \cite{dePater2019}. We define the base of the water cloud ($\sim$ 5 bar), sitting just above the Lifting Condensation Level (LCL) of water around $\sim$ 6-7 bar.  The build-up of internal energy in the form of Convective Available Potential Energy (CAPE) as the source for these outbreaks  is consistent with our understanding of giant storms on Saturn \cite{Li2015}. Furthermore these outbreaks are part of the SEB revival cycle after it fades \cite{Fletcher2017} and hence recurring phenomena. Observations at optical wavelengths of the storm system in March 2017 found that the storm clouds were more extended in depth compared to the typical cloud deck in the surrounding regions, consistent with a deep source for this region \cite{dahl2021}. Observations at 5 $\mu$m showed an increase in ammonia and water abundance coupled with an opaque water cloud at the location of the storm plume, confirming that the plume is bringing up water and ammonia from deeper levels within the atmosphere \cite{Bjoraker2022}. 

\section*{Results}\label{sec:R}
%Data 
We use a combination of Juno MicroWave Radiometer (MWR) and Hubble Space Telescope (HST) observations to constrain the vertical extent of the storm system. The MWR data were obtained during PJ4 (Feb. 2nd, 2017, closest approach at UTC 12:57), when Juno flew over the still ongoing storm region and observed the atmosphere at six distinct frequencies in the radio regime \cite{janssen2017}. The signal is the opacity-weighted brightness temperature of the atmosphere integrated over a range of pressures, with ammonia being the dominant opacity source in the microwave on Jupiter; however, the observations are also sensitive to physical temperature variations. The horizontal resolution of the observations is set by the convolution of the atmospheric thermal emission with the sensitivity pattern of the beam and thus is variable across the domain. The vertical resolution, that is the depths over which the thermal signal is integrated, is frequency dependent. At lower frequencies, the observations probe deeper into the planet and also cover a wider pressure range (see fig. \ref{fig:WF}). From the calibrated antenna temperatures, we retrieve the nadir brightness temperature and limb-darkening parameters as a function of latitude \cite{moeckel2023}, and fit vertical atmospheric profiles to these quantities using a radiative-transfer model (radioBEAR - Radio BErkeley Atmospheric Radiative-transfer (\url{https://github.com/david-deboer/radiobear} \cite{dePater2005,dePater2014,dePater2019b,moeckel2023}). To resolve smaller-scale structures in the brightness-temperature distribution, we use a deconvolution algorithm to map radio observations onto spatially resolved HST storm observations (see Methods). \\ 
The HST observations were taken in support of the Juno mission during Juno's flyby (PI: Mike Wong). The UVIS detector on the WFC3 instrument covers the UV, visible, and near-infrared in seven distinct filters \cite{dressel2023}. The data were processed and navigated \cite{wong2020}, and are available on the Barbara A. Mikulski Archive for Space Telescopes (\url{https://archive.stsci.edu/hlsp/wfcj}).  

\begin{figure}[h]
    \centering
    \includegraphics[width=\textwidth]{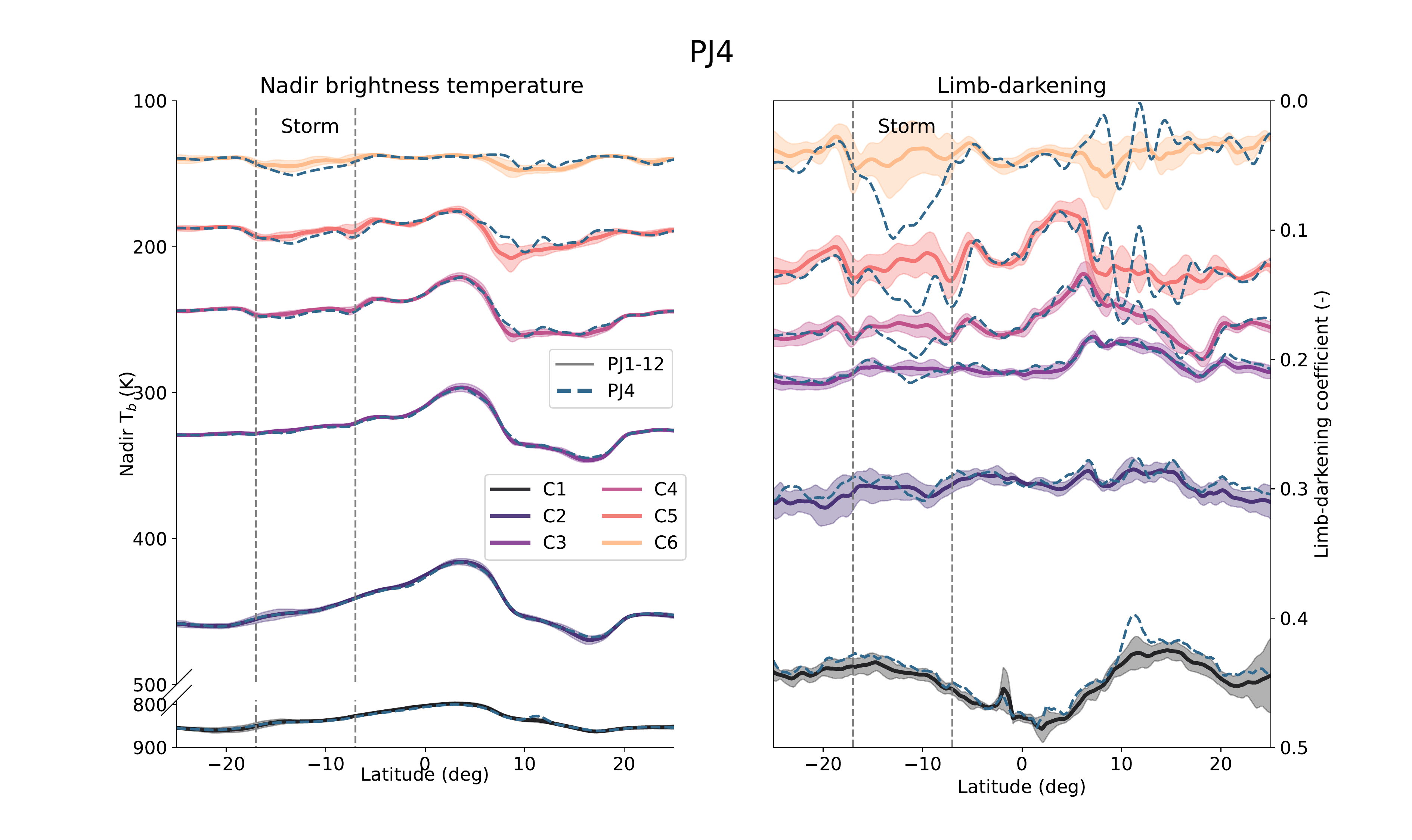}
    \caption{\textbf{Comparison of PJ4 observables (dashed lines) with the mean observations (solid lines, shading corresponds to the standard deviation) based on the first 12 perijoves.} The nadir brightness temperature is on the left, while the limb-darkening coefficient is shown on the right with each color corresponding to a different frequency for a given channel. The effect is strongest for the upper channels probing the upper atmosphere, with its strength decreasing at lower frequencies. Channel 5 and 6 probe the pressures around the ammonia ice clouds. Channel 3 and 4 probe pressures near and below the water cloud, while Channel 1 and 2 probe deepest into the planet - see fig. \ref{fig:WF}. We see an increase in brightness temperature at the location of the storm, indicating either a drying out of the upper atmosphere in ammonia or an increase of the physical temperature. } 
    \label{fig:PJ4obs}
    \end{figure}

% Results
Even about a month after the initial outbreak, the longitude-averaged observations shown in Fig. \ref{fig:PJ4obs} show a distinct signature in the SEB at the location of the storm outbreak. The majority of the storm signal is confined to the upper 4 channels, as seen in both the nadir brightness temperature and the limb-darkening coefficient (signal exceeds the standard deviation in Channel 3-6) when comparing the observations with the mean of the first 12 perijoves \cite{moeckel2023}. The nadir brightness temperature $T_{B0}$ is the equivalent radio-temperature that matches the observed flux for nadir pointing ($\theta = 0$), and the limb darkening $p$ is the reduction of the flux at a given slant angle $\theta$  $T_b(\theta) = T_{B0}(\theta)cos(\theta)^p $ following the convention by \cite{sault2004}.

Channel 3 probes pressures greater than the expected water condensation pressures and thus probes below the traditional weather layer. The fact that we can trace this signal all the way to Channel 3 already indicates that this storm is not contained to the traditional weather layer (cloud condensation region, P$<$6bar), where we see the majority of variability in the atmosphere \cite{dePater2016}, but affects the atmosphere to much greater pressures.

\begin{figure}
    \centering
    \includegraphics[width=\textwidth]{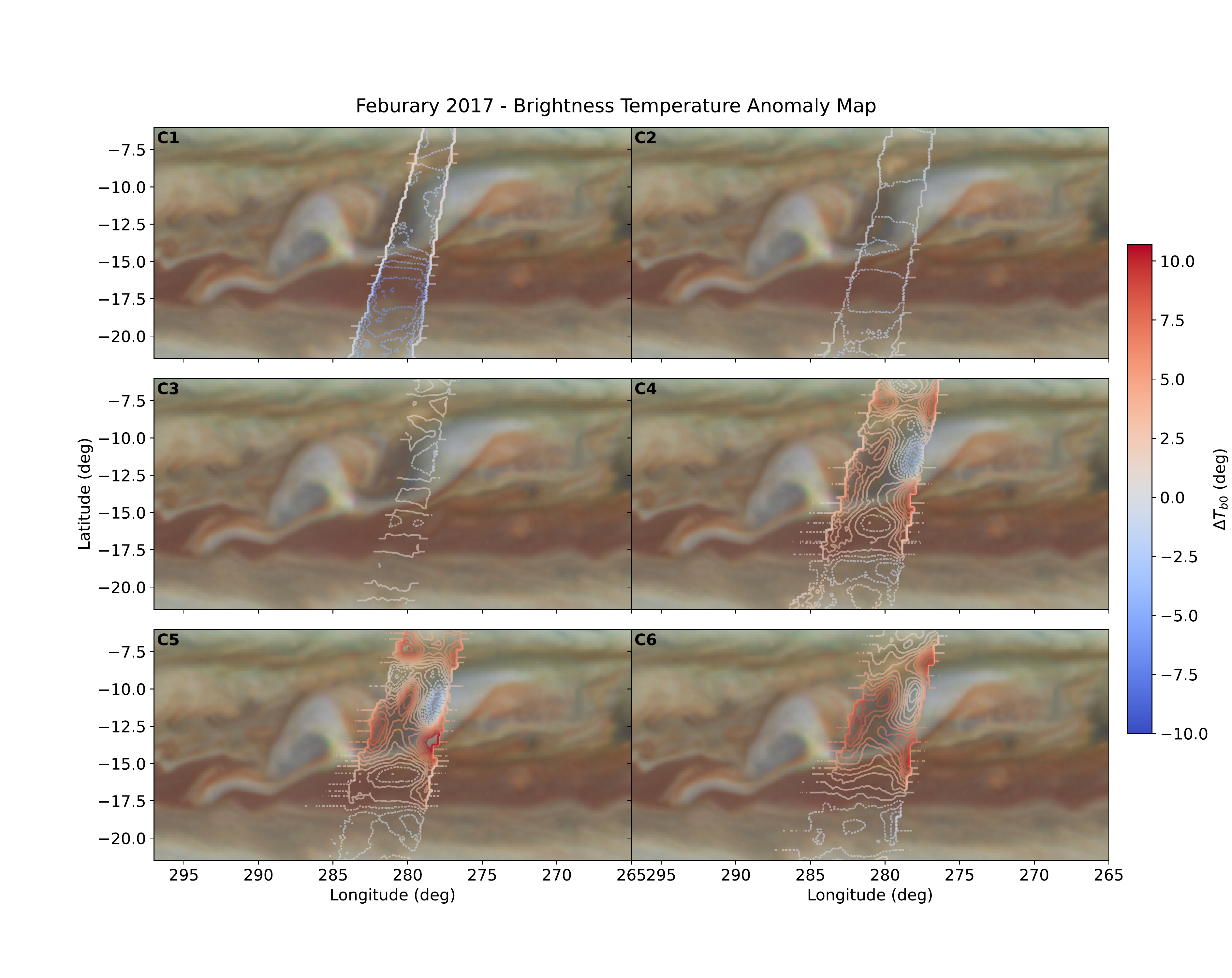}
    \caption{\textbf{Brightness temperature anomaly maps for each channel.} The contours outline the difference between the mean background atmosphere (PJ1-PJ12) and PJ4. The positive outlines indicate an atmosphere that is radio-warmer than the average atmosphere, while blue contours indicate a radio-colder atmosphere. All six maps are shown on the same color-scale showing that the signal is strongest in the C6 and C5 channels, which sound pressures less than 3 bar (see fig. \ref{fig:WF}). The signal strength decreases progressively from C4 ($\sim$3 bar) to C1 ($\sim$40 bar).}
    \label{fig:PJ4maps}
    \end{figure}

The deconvolved brightness temperature maps detail the finer-scale structure within the storm system, highlighting the variations across the storm domain. The deconvolved measurements in Fig. \ref{fig:PJ4maps} depict the brightness-temperature anomaly overplotted on concurrent HST observations. We define the anomaly as the difference in the PJ4 nadir brightness temperature for a given channel and the mean based on the first 12 orbits. Higher frequencies (C5 and C6) probe the region where the ammonia ice clouds are forming, and the cold brightness temperature correlates well with the haze and cloud morphology as captured by the HST observations. The cloud itself is surrounded by visually darker regions, associated with a depleted ammonia abundance \cite{Bjoraker2022}, which is confirmed by a warm brightness temperature anomaly. The maps indicate that the brightness temperature anomaly decreases at longer wavelengths (C3 and C4) but remains co-located with the storm topology. At the lowest frequencies (C1 and C2), which probe deepest into the planet, the brightness temperature anomaly becomes negative across the domain. The two lowest channels have the widest beam-width and the least amount of coverage due to filtering of synchrotron radiation. Thus the results in C1 and C2 should be seen as an upper limit, that is the negative brightness residual might be even colder. The location of the previously studied upwelling plume \cite{Bjoraker2022} is slightly west of the covered domain, with our observations covering an adjacent plume.

\begin{figure}
    \centering
    \includegraphics[width=1.0\textwidth]{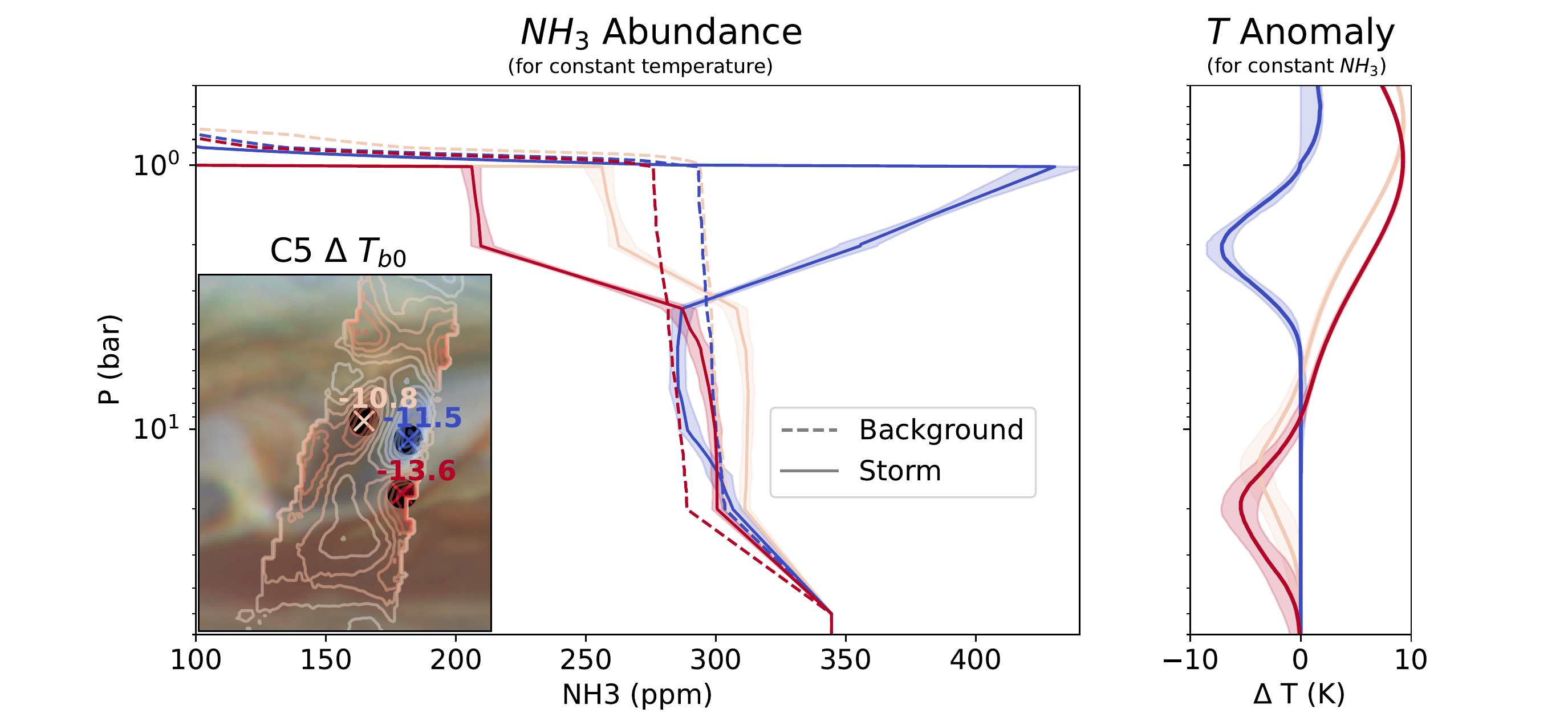}
    \caption{\textbf{Retrieved ammonia abundances that show the impact of the storm on the atmosphere.} Left panel: Vertical ammonia abundance profiles that compare the mean background atmosphere (dashed line) with the results from the relative retrieval assuming no changes to the physical temperature structure (solid lines). The locations of the profiles are indicated on the inset map that shows the C5 brightness temperature anomaly. The shading corresponds to the uncertainty in the retrieved solutions. For all three profiles, we find a strong anomaly in the upper atmosphere, sitting over a smaller, but opposite sign anomaly located below the water cloud. Right panel: Retrieved temperature anomaly with the ammonia abundance fixed at the background level as shown by the dashed line in the left panel. The regions adjacent to the clouds are characterized by strong atmospheric heating in conjunction with cooling deeper in the atmosphere. }
    \label{fig:RelRetrieval}
    \end{figure}

We selected three regions (10.8S, 11.5S, and 13.6S) in Fig. \ref{fig:PJ4maps} to retrieve the ammonia abundance to study the effect of the storm on the atmosphere. In Fig. \ref{fig:RelRetrieval} we show the impact of the storm on the atmosphere of Jupiter, either in the form of disrupting the ammonia profile or by affecting the temperature structure. We first fit a reference profile (dashed lines) based on the averaged quantities from the first 12 perijoves at those latitudes coupled to a dry adiabat. Relative to this background profile, we then find ammonia abundance solutions (solid line + shading) that reproduce the observed brightness temperature anomaly shown in Fig. \ref{fig:PJ4maps} and fig. \ref{fig:AnomalyFit} within the given observational constraints. The colors correspond to the locations on the inset map: the blue color is centered on the storm plumes with the largest negative brightness temperature anomaly; the dark red color corresponds to the southern edge of the clouds where we find the strongest positive anomaly; and yellow corresponds to the region just west of the cloud. For now, we assume that there is no change in the temperature structure and we attribute the brightness temperature anomaly fully to the ammonia abundance, rendering this result an upper limit to the ammonia variability. 
The profiles show two distinct regions, with the majority of the variability occurring at altitudes above the 4 bar level, as shown by the strong signal in the upper three channels. The two warm regions are characterized by a depletion of ammonia, while the radio-cold regions show a large localized ammonia enhancement that peaks around the ammonia ice condensation pressure. Surprisingly, deeper in the atmosphere, the ammonia anomalies change sign. In the region where the gas is enhanced at the ammonia cloud deck, we require a depletion that extends down from the water cloud to $\sim$ 15 bar to fit the observations. In the regions where we find a depletion in the upper layers, we find a small but significant increase in the ammonia abundance that starts around 4 bar and extends to $\geq$ 20 bar. 

% Explain results for Temperature anomaly here
In addition to affecting the distribution of ammonia, the outbreak is expected to also significantly heat the atmosphere through the release of latent energy as trace gases condense in the upward branches of the storm and through subsidence of air in a stably stratified atmosphere. To explore the other end-member case, we kept the ammonia abundance constant at the averaged background level, and instead we fit a pure physical temperature anomaly. We parameterize the temperature anomaly with two Gaussian temperature deviations from the nominal dry adiabatic profile, where each Gaussian has an associated depth, width, and magnitude. The retrieved deviations in temperature, along with their uncertainties, are shown on the right side of Fig. \ref{fig:RelRetrieval}. We find that the temperature anomaly is inversely proportional to the ammonia anomaly. Instead of a depletion in ammonia, the observations are also consistent with strong heating in the upper atmosphere, with the retrieved physical temperature anomaly similar in magnitude to the observed brightness temperature anomaly. Below the water cloud base, the observations may indicate significant cooling at pressures between 10-20 bar. At these higher pressures, the required physical temperature anomaly that fits the observations is a few times larger than the observed brightness temperature anomaly. 
The ammonia enhancement in the storm plume could also be explained by a strong cooling at 1-4 bar, with the physical temperature anomaly and the brightness temperature anomaly at roughly the same magnitude. In this scenario, no further change in the physical temperature below the plume was required to fit the observations. Comparing the observed brightness temperature anomaly with the retrieved brightness temperature anomaly shows that both assumptions explain the observations well, despite very different structures in the atmosphere (see fig. \ref{fig:AnomalyFit}).

We attempted to break the degeneracy by allowing both the temperature and ammonia to vary simultaneously across the full pressure domain. With the available measurements, we found no significant constraint on either of them. Note that we do not have an independent measurement for the limb-darkening effect for our analysis.

\begin{figure}
    \centering
    \includegraphics[width=0.75\textwidth]{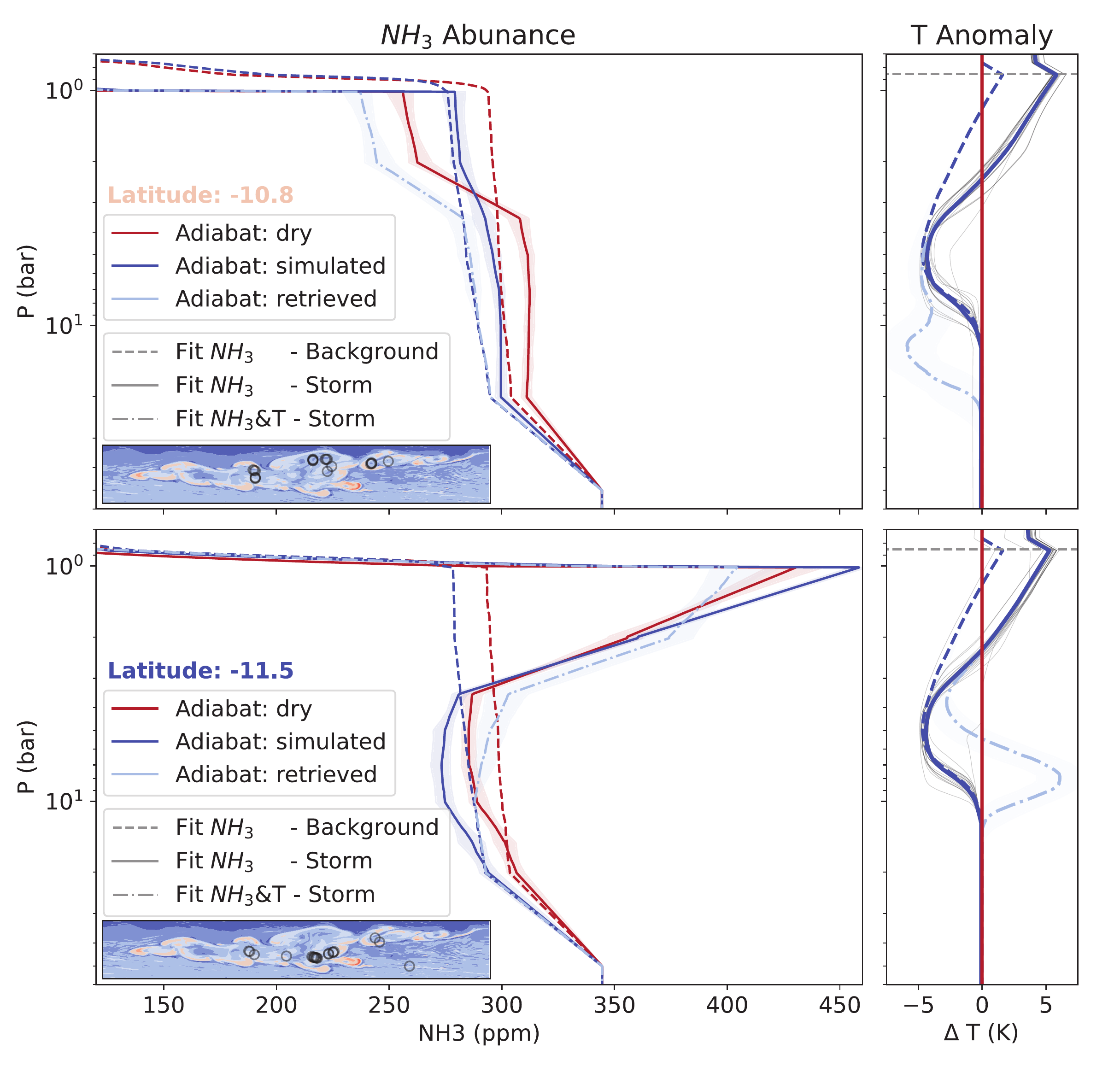}
    \caption{\textbf{Retrieved ammonia abundance results for different assumptions about the physical temperature structure in the atmosphere.} The upper panel corresponds to the depleted region west of the storm plume (10.8$^\circ$S on inset Fig. \ref{fig:RelRetrieval}), while the lower panel shows results for the region that is co-located with the storm plume (11.5$^\circ$S). The red lines indicate the background (dashed) and storm (solid) profile for our nominal assumptions of a dry adiabat, while the blue lines show the results including physical temperature variations, as shown on the right. We have included a superadiabatic layer above the water cloud (dashed line), a subadiabatic region during the storm outbreak (solid line), and a Gaussian temperature anomaly that fits the data assuming that there is no ammonia variation below 10 bar (dot-dashed line). The thin gray lines show a range of temperature anomalies that we obtained from the simulation, where each location is indicated by the gray circle in the inset map on the right-hand side.} 
    \label{fig:RetrievalPanel}
    \end{figure}

The above analysis represent limit cases. To assure that our results are robust, we use simulated temperature structures based on non-hydrostatic simulations \cite{Li2019} that allow us to both compare the retrieved temperature structure to the simulated values, and also to study the correlation between ammonia and temperature anomaly. 
In Fig. \ref{fig:RetrievalPanel} we fit the anomalies at 10.8 $^\circ$ S (upper panels) and 11.5 $^\circ$ S (lower panels) and retrieve the abundances of ammonia for a range of different simulated temperature profiles. The right-hand side of Fig. \ref{fig:RetrievalPanel} shows the temperature deviation with respect to a dry adiabat informed by the simulations, where each line style corresponds to a different assumption. The inset map shows a cross section of the simulated storm outbreak that shows the ammonia abundance around 1 bar: regions in red are depleted in ammonia, while blue corresponds to enhanced ammonia. %Our aim is to show that a more realistic solution informed by the simulated temperature structure lies between the two limits shown in Figure \ref{fig:RelRetrieval}. 

%Background atmosphere 
The averaged temperature structure in the simulated domain is a good proxy for the mean atmospheric temperature that we use to fit the background profile (blue dashed lines). The most significant feature is a superadiabatic layer around and above the lifting condensation level of water due to the mass loading of water, which is heavier than the predominantly hydrogen atmosphere \cite{Guillot1995,Sugiyama2014,Leconte2017,Li2018,Li2019}. The tentative superadiabaticity is supported by a recent revisit to the MWR retrieval of the background atmosphere and regional numerical simulations \cite{Li2024}. Comparing the blue and red dashed lines, we find that including superadiabatic temperature structure in the weather layer of Jupiter results in less ammonia to fit the data. Interestingly, while the superadiabatic temperature layer is limited to altitudes above 10 bar, the effects on the retrieved ammonia abundance can be traced to much greater depths due to the broad weighting functions (fig. \ref{fig:WF}). 

% Storm atmosphere 
Storms occur periodically every three to four years in the simulation, and we choose a snapshot of the simulation during an on-going outbreak. The simulations allow us to compute local temperature anomalies, that is the change of temperature with respect to the simulated background reference profiles. For our storm profiles (solid blue lines), we averaged 20 distinct temperature anomalies (thin gray lines) in the simulation, where each profile was selected based on a L2-regularization that minimizes the difference between the retrieved profile in Fig. \ref{fig:RelRetrieval} and simulated ammonia profiles. The small inset map on the left-hand side is a cross-cut through the simulated ammonia abundance at 1 bar and the 20 dots show that we sampled a number of distinct regions in the simulations. In addition to the superadiabatic layer, the storm profiles predict strong heating of the upper atmosphere as a consequence of the latent heat release of water. For regions with a positive brightness temperature anomaly (upper row), our observations of heating in the upper atmosphere are consistent with the simulations. Indeed, the simulated temperature increase in the upper atmosphere is large enough to explain the signal in the upper channels and we do not require any depletion in ammonia above 4 bar (blue solid line). At higher pressures, we still require an anomaly in the ammonia abundance that extends down to $\sim$ 20 bar. For the storm plume (lower row), the region where we found the strongest cold brightness temperature anomaly, we require a slight increase in ammonia abundance to compensate for the warmer upper troposphere.

In reality, the brightness temperature anomaly is most likely caused by a combination of ammonia and thermal anomaly. We leverage the simulation results, which show no significant variation in ammonia abundance below 10 bar, and restrict the ammonia anomaly to only occur at altitudes above 10 bar. In addition to the ammonia anomaly, we introduce a Gaussian temperature deviation. The upper row shows that we could fit the region adjacent to the cloud through a combination of a depleted upper atmosphere down to 4 bar, sitting over a cold-temperature anomaly around 12 bar (dot-dashed cyan lines), at a pressure similar to that obtained for the pure temperature anomaly (see Fig. \ref{fig:RelRetrieval}). For the retrieval of the storm plume, we find a persistent but slightly decreased ammonia enhancement under the clouds (left-hand side), with a strong positive physical temperature anomaly concentrated around 7.5 bar.

\section*{Discussion}\label{sec:D}
Previous observations \cite{Gierasch2000,Fletcher2017,dePater2019,dahl2021,Bjoraker2022}, simulation results \cite{Li2019} and results from comparable storms in the solar system \cite{Porco2005,SanchezLavega2008,Li2015} all indicate that water is the main carrier of the energy once the large-scale storm breaks after having accumulated sufficient energy to overcome the convective inhibition. These are the first observations to quantify the effect of these storms on the ammonia abundance and temperature structure below the visible cloud deck and show how deep the atmosphere is disrupted in the aftermath of the storm. Our retrievals reveal two distinct regimes that have been affected by the storms: a strong anomaly located above 4 bar and a smaller, but opposite anomaly, extending well below the water cloud base to about 20 bar. We have fit several different ammonia and temperature profiles to the brightness temperature anomalies and found that a wide range of models is consistent with the brightness temperatures. 

The visually darker regions are indicative of a lack of thick ammonia clouds and are interpreted as a sign of down-welling of dry air. Our observations are consistent with subsidence regardless of the assumptions about the main source for the observed anomalies. We observe subsidence in the form of depleted air \cite{Showman2005,Sugiyama2014} or adiabatic heating in a stably stratified atmosphere \cite{Ingersoll2004}. The heating of the upper atmosphere is also consistent with non-hydrostatic simulations that predict a subadiabatic region from 4 bar upwards. 

At higher pressures below the visually dark regions, the observations point towards processes that transport condensates well below the level where the storms erupted from, causing either an abundance or temperature anomaly below the water cloud. For the former case, the retrieved ammonia anomaly switches signs around 4 bar, consistent with the maximum depths that large ammonia rain droplets could reach \cite{Li2019}. The fact that we find a persistent enhancement at much higher pressures requires processes that deliver ammonia to temperatures well above its boiling point and below the predicted depths from simulations that  include standard droplets \cite{Sugiyama2014,Loftus2021}.

Alternatively, if we assume that the signal has a thermal anomaly contribution, we would require significant cooling around 15 - 20 bar if it is purely thermal (see Fig. \ref{fig:RelRetrieval} or more shallow around 5 - 15 bar assuming a combined ammonia-temperature anomaly (see Fig. \ref{fig:RetrievalPanel}). The temperature anomaly is located about one scale height (H) below the water condensation pressure and with that deeper than the maximum depth (0.5H) predicted for the largest stable water droplets to fall under Jovian conditions \cite{Loftus2021}, but still above the boiling point of water at $\sim$ 30 bar \cite{Li2019}. Instead, we argue that the depth of the anomaly requires even larger precipitates, consistent with the so-called mushballs, a mixture of frozen water and ammonia ice that can form in regions of strong updrafts \cite{Guillot2020}. Both the depths of the anomaly ($\sim$15 bar) and the observed temperature anomaly ($\sim$4 K) are consistent with the predicted range for pressure (5 - 30 bar) and temperature (2 - 17 K at 15 bar).  We note that the mushballs could also cause an ammonia anomaly at these depths, and indeed simulations of the processes indicate that they should affect the atmosphere down to pressures of $\sim$ 30 bar \cite{Markham2023}. 
   
In contrast, for the storm plume at 11.5$^{\circ}$S we find that the ammonia anomaly is consistent with the storm dredging up ammonia and thereby drying out the deeper atmosphere and enhancing the upper atmosphere. The increase of ammonia  from 3 bar until the cloud base around 1 bar follows the signature of local condensation and evaporation around the cloud top \cite{Weidenschilling1973,Ingersoll2017,Li2020}. While the cooling around 2 bar is consistent with ammonia evaporation, it is unlikely that ammonia can cause such a strong temperature anomaly. Neither the expected temperature change on the order of 1 Kelvin \cite{dePater1986} nor the simulations are consistent with such strong cooling. Therefore, we prefer ammonia as the main source for the anomaly higher up in the atmosphere. Indeed, the in-between case, where ammonia and temperature are allowed to vary (dot-dashed line in Fig. \ref{fig:RetrievalPanel}), reconciles both ammonia and temperature anomaly with our expectations for up-welling within the storm plume. In the ascending branches of the storm, a significant amount of latent heat is released as the water condenses. The heating at 7.5 bar points to significant condensation of water, consistent with the existence of the opaque water cloud in these storm plumes \cite{Bjoraker2022}. The latent heat release further fuels the ascent of the parcels, delivering ammonia to the top of the atmosphere forming thick ammonia ice clouds \cite{dahl2021,Bjoraker2022}. The combined ammonia and temperature interpretation confirms the role that moist convection plays in providing energy for the storm to maintain itself for long periods \cite{Ingersoll2000,Gierasch2000,Li2015}, and an excellent indicator of how dynamics and microphysics could interact to shape the atmosphere of Jupiter, and possibly on the other giant planets. 

Although our results are encouraging, there are a few open questions. The retrievals for the regions surrounding the clouds require the upper atmosphere to be subsiding, while the enhancement in ammonia or cooling below the water cloud necessitates rain-out and evaporation. Could the subsidence of the upper atmosphere due to the outbreak trigger secondary smaller-scale storms that produce mushballs? 
In the region where the upper atmosphere is consistent with strong updrafts as inferred from both in the HST images and from the radio retrievals, the lower atmosphere appears either depleted in ammonia gas or significantly heated, contrary to what moist downdrafts \cite{Showman2005} or the mushball evaporation model predict \cite{Guillot2020}.

Regardless of the interpretation of the observations, we find that storms originating at the water condensation pressure have a significant impact on the atmosphere of Jupiter. This is the first time that we see observationally how the ammonia and water cycles are coupled, and how storms redistribute the energy and trace gases in the atmosphere. These outbreaks are often long-lived, lasting for many months, and we show how they continuously disrupt the atmosphere of Jupiter as they evolve. This confirms storms’ crucial role in heat transport within the atmosphere \cite{Gierasch2000}, their capacity to alter the appearance of entire belts, \cite{Fletcher2017,dePater2023} and help explain the global depletion of ammonia down to $\sim$ 30 bar as determined by MWR radio observations \cite{li2017,moeckel2023}.

\section*{Methods}\label{sec:methods}
\subsection*{Deconvolution Algorithm}\label{ssec:MDA}
We developed an algorithm that obtains sub-beam size brightness temperature distributions based on overlapping observations. This method was developed for use with Juno's MicroWave Radiometer (MWR) due to its irregular sampling of the domain during the flyby, but it is generalizable to any type of observation. The algorithm compares each individual observation with a best estimate of the brightness temperature on the planet, and the differences are then remapped onto the planet. This approach is similar to the approach that retrieves the gridded distribution of the brightness temperature \cite{Zhang2020}. The final resolution of the map is determined by the overlap of the individual observations, resulting in non-uniform resolution throughout the domain.  

To obtain a first estimate of the brightness distribution on the planet, we bin all observations at a given frequency $\nu$ as a function of boresight latitude $\theta$ \cite{moeckel2023}. The zonal nadir brightness temperature $T_{b0}(\theta,\nu)$ and the limb-darkening coefficient $p(\theta,\nu)$ are then obtained by fitting a parametric relationship between the observed brightness temperature and the emission angle \cite{moeckel2023}. In fig. \ref{fig:C5Footprints} we show the C5 observations color-coded by the nadir brightness temperature to show the longitudinal coverage obtained for a given flyby geometry. The slanted bar to the east represents the fitted zonal average nadir brightness temperature $T_{b0}(\theta,\nu)$ corresponding to the colorbar on the right hand side.

We aim to construct a two-dimensional map of the nadir brightness temperature on the planet $T_{b0}(\theta,\phi,\nu)$ that matches the observations. For each integration, we compute the transformation from the instrument frame (azimuth and elevation) of each channel to the planet frame (longitude and latitude): ($(\theta_{sc},\phi_{sc})$ $\rightarrow$ $(\theta,\phi)$) and map the beam pattern onto the planet out to twice the half-power beam width \cite{moeckel2023}. For each facet of the beam we obtain the corresponding location on the planet and the emission angle for each beam element with respect to the spacecraft ($\mu \equiv cos(\theta_e)$). By convolving the brightness temperature map with the projected beam sensitivity and the emission angle distribution across the planet, we can then simulate the observed antenna temperature $T_A^{sim}$: 

\begin{equation}
    T_A^{sim}  = \frac{\int \int_{0}^{\theta}G(\theta,\phi)T_b(\theta,\phi)\mu^{p(\theta,\phi)}sin(\theta_{sc}) d\theta_{sc}d\phi_{sc}}{\int \int_{0}^{\theta}G(\theta,\phi)sin(\theta_{sc}) d\theta_{sc}d\phi_{sc}}
    \label{eq:Taest}
\end{equation}

%Need a figure to show the difference 
The mismatch between the expected antenna temperature and the actual observations: $\Delta T_A = T_A^{sim} - T_A^{obs}$ corresponds to a missing structure on the scale of the beam within the map. We can then update the brightness temperature map for each observation with the computed mismatch: 

\begin{equation}\label{eq:Tb0i+1}
    \Delta T_{b0}(\theta,\phi) =  \Delta T_A G(\theta,\phi),
\end{equation}
where the beam ($G$) is normalized so that the residual integrated over the beam is $\Delta T_A$. 

For the zonal average, we obtained an independent measure of the limb-darkening coefficient by comparing measurements at different emission angles for a given latitude. By convolving the nadir brightness temperature map with the observation geometry, we can no longer directly constrain the limb-darkening effect. Instead, we must assume prior knowledge for the limb-darkening coefficient to compute the simulated antenna temperature. The simplest assumption ignores the longitudinal variation in the limb-darkening coefficient and uses the zonal average \cite{Zhang2020}. If we interpret variations in the brightness temperature as variations in the atmosphere, it is expected that the limb-darkening coefficient changes accordingly with longitude. Therefore, we calculated the local correlation between the nadir brightness temperature and the limb-darkening coefficient ($\partial p/ \partial T_{b0}$) from the zonally averaged data and linearize the system to obtain an estimate for the variation in the limb-darkening coefficient. 

\begin{equation}
\Delta p = \frac{\partial p}{\partial T_{b0}} \Delta T
\end{equation} 
% Our brightness distribution corresponds to the nadir brightness temperature $T_{b0}$, while the residual is the beam-convolved brightness distribution modulated by the emission angle effect $\mu$: $\int G T_{b0} \mu$. 
The residual that we obtain is limb-darkened, however, we do not correct for the emission angle effect when we are adding back the residual to the map. If we define the beam-integrated emission effect to be $\bar{\mu}$, this effectively reduces the amplitude of the residual by the factor $\bar{\mu}$. While this approach slows down the convergence of the residual, it has the advantage that this method is less sensitive to the assumption about the limb-darkening effect and reduces spurious artifacts for beam facets with large emission angles.   

We repeat this process for each individual observation and sum up each individual residual to obtain a two-dimensional distribution of brightness temperature and a correlated limb-darkening coefficient. We also keep track of the number of observations (contribution map) constraining a given region using a normalized projected beam, where the center of the projected beam is scaled to unity. 

After all observations have been processed, we normalize the system by dividing the $\Delta T_{b0}(\theta,\phi)$ - map with the contribution map. Panel A) of fig. \ref{fig:PJ4-DTMap} shows an example of the $\Delta T_{b0}$ - map for Channel 5, which shows the summed residual after convergence is reached (15 iterations). The cross section of panel B) indicates how much structure is added during each iteration, with the added signal decreasing with each iteration. The y-axis on the right shows a cross-cut through the contribution map, a proxy for the number of measurements for a given region. Towards the center of the domain, we have values of up to 75, with the distribution falling off rapidly towards the edges resulting in a roughly 10 degree slice.  To avoid edge effects caused by poor sampling toward the boundary of the sampling domain, we have chosen to taper the residuals for regions where the contribution map falls below a value of 5, disregarding a significant number of observations. 
The shape of the retrieved residuals correlates well with the cloud and haze morphology of the concurrent HST observations, affirming that we are obtaining realistic longitudinal variations. We then add the $\Delta T$ - map and $\Delta p$ - map to our previous estimate and start the next iteration.

We track the convergence of the method by comparing the standard deviation of the retrieved mismatch with the theoretical error based on MWR error budget \cite{janssen2017} (see fig. \ref{fig:Convergence}). The error budget specifies 0.1\% of the individual antenna measurement, which translates to $~$0.2K in Channel 5. The method converges within 5-15 iterations for most observations, upon which further iterations no longer improve the overall mismatch. The remaining mismatch between observations and the model can be explained by the effect of limb-darkening variation across longitude or structure on scales smaller than we can resolve.  The final product is a distribution of nadir brightness temperature with their resolution set by the observation geometry: close to the center, the map has the highest resolution, which decreases towards the edge due to larger individual footprints and less overlapping observations.

To decrease the impact of synchrotron radiation at the longer wavelengths, we removed observations when the beam is pointing toward the equator \cite{moeckel2023}, where the strongest source of synchrotron radiation is expected to be in the sidelobes \cite{SantosCosta2017}. This reduces the available measurements for the deconvolution algorithm for C1 - C3 and limits the width of the final maps at the lowest frequencies, as seen in Fig. \ref{fig:PJ4maps}.

\subsection*{Relative Retrieval}\label{ssec:MRR}
We are interested in the change in the atmosphere due to the ongoing storm. Therefore, we aim to retrieve a brightness temperature anomaly compared to our mean model that matches the observed brightness temperature anomaly for the PJ4 observations. This represents a relative measurement, which reduces the impact of the absolute calibration uncertainty and puts tighter constraints on the effect of the on-going storm system. This comes with the caveat that our retrieved abundances are only valid with respect to the mean atmospheric model.

We first identify the latitude of regions with large anomalies in Fig. \ref{fig:PJ4maps}, for which we retrieve the background atmosphere. For the mean background atmosphere we use the zonally averaged brightness temperature ($T_b$ observed brightness temperature, $T_m$ modeled brighntess temperature at a given frequency represented by the 6 channels) and limb-darkening coefficients ($p$ observed, $p_m$ modeled) with their corresponding uncertainties ($\sigma_T$, $\sigma_p$, respectively) to constrain the vertical distribution of ammonia abundance \cite{moeckel2023}. In practice, we  minimize the log likelihood of the following marginalized probability distribution $\Tilde{p}$  using a Markov Chain Monte Carlo (MCMC) framework:  

\begin{equation}
    \begin{split} 
    ln \left( \Tilde{p}(T,p\rvert x,\sigma) \right) =& -0.5\left(\sum_{1}^{6}  ln(2\pi\sigma_{T}) + \frac{(T_{m} - T_b)^{2}}{\sigma_T^{2}} +  \sum_{2}^{5}  ln(2\pi\sigma_p) \frac{(p_{m} - p)^{2}}{\sigma_{p}^{2}} \right)  \\ 
    &+ \lambda \mid \sum_{1}^{8}{\left(\frac{\delta NH_3}{\delta ln(P)}\right)_i} \mid
    \end{split} 
    \label{eq:CF_mean}
\end{equation}

We are interested in the consequence of the dynamical impact of the storm, and we chose a regularized stochastic model where the pressure can change between preset pressures \cite{li2017,moeckel2023}. We have selected our nine nodes to sample the atmosphere around the ammonia cloud and the water cloud pressure to constrain the dynamical impact of trace gas condensation ($P_{nodes}$ = 1, 2, 3.5, 5, 7, 10, 15, 20, 50 bar). The regularization, the last term on the right hand side of Equation \ref{eq:CF_mean}, aims to reduce the amount of vertical fluctuations by minimizing the change in ammonia with pressure ($\lambda$ = 1E4 \cite{moeckel2023}). Based on the average of the last 5000 draws from the MCMC chain, we can model the brightness temperature $\bar{T}_{model}$ corresponding to a given latitude. 

Once we have obtained the mean background model, we can retrieve an ammonia distribution that produces the same brightness temperature anomaly $\Delta T_{model}$ as observed during PJ4 $\Delta T_{observation}$. We use the same pressure nodes as we have done for the background atmosphere. For the error $\sigma_{\bar{T}}$, we use the $0.1\%$ of the observed brightness temperature \cite{janssen2017}. We minimize this cost function:  

\begin{equation}
    \begin{split} 
    \ln \left( p(T\rvert x,\sigma) \right)  = &-0.5\left(\sum_{i=1}^{6}  \ln(2\pi\sigma_{\bar{T}}) + \frac{\overbrace{(T_{storm} - \bar{T})^{2}}^{Model} -  \overbrace{(T_{storm} - \bar{T})^{2}}^{Observations}}{\sigma_{\bar{T}}^{2}} \right) \\ 
    &+ \lambda  \left| \sum_{i=1}^{8}\left({ \frac{\delta NH_3}{\delta \ln(P)}}\right)_i \right|
    \end{split} 
    \label{eq:CF_rel}
\end{equation}

Our final results are the ammonia abundance anomaly or physical temperature anomalies in the form of a Gaussian temperature deviation, which matches the observed brightness temperature anomaly as shown in Fig. \ref{fig:RelRetrieval}. We parameterize the temperature anomaly as a Gaussian in the log pressure space, where we fit the depth, width, and magnitude, introducing three more variables to fit. 

\subsection*{Temperature Structure}\label{ssec:TS}
The observed brightness temperature variations can be ascribed to both variations in the trace gas abundances and variations in the physical temperature as a consequence of the ongoing storm complex. We tested this correlation by assuming various temperature profiles and determine its impact on the retrieved ammonia abundance. 

Our nominal assumption is that everything follows a dry adiabat anchored to the 1-bar measurement of the Galileo probe (166.1K $\pm$ 0.8) \cite{Seiff1996}. This temperature profile is a good representation when studying large-scale dynamics, as evaporation and condensation of trace gases are expected to act on much smaller scales and the result of the Galileo probe showed that within the 5 $\mu$m hot spot the atmosphere was close to a dry adiabat.

To quantify the impact of the storm on the temperature distribution, we employ non-hydrostatic simulations for Jupiter \cite{Li2019}, which include the effect of condensable species on the dynamics of the atmosphere. The simulations captured a large-scale storm system at mid-latitudes, allowing us to capture the correlation between ammonia abundance and temperature variations. Based on the simulations we construct two temperature profiles: a domain-averaged temperature structure that is used to constrain the background atmosphere and individual profiles that represent the physical temperature variation during an ongoing storm. 

For our updated background profile, we let everything at pressures below 12 bar follow our nominal dry adiabat, making 12 bar the anchor point where both profiles agree. The temperature structure below is highly dependent on the water content and could be significantly warmer than our assumptions \cite{Leconte2017}, however, this effect would only affect the absolute ammonia abundance and has less impact on our relative retrieval. 
At altitudes above 12 bar, the temperature profile is affected by the stabilizing effect of the water cloud, causing a superadiabicity around the water cloud condensation pressures. We compute the adiabatic lapse rate above 12 bar based on the simulations and propagated it upward until the 1-bar pressure. At 1-bar, the temperature reaches 166.9K, within the errors of the Galileo probe measurement \cite{Seiff1996} and within a degree from our nominal profile. Higher in the atmosphere, the temperature structure follows the results from mid-infrared observations \cite{Fletcher2009}. 

For the individual storm profiles, we match the retrieved ammonia profiles with the simulations that exhibit a similar topology. For each region, we selected 20 profiles from the simulated domain and computed the mean temperature deviation. All profiles share a strong subadiabatic region above 4 bar, leading to local heating of the troposphere. 

% % Temperature pulse results, flip the script
% In the above methodology, we ascribe all variability that we see at pressures greater than 10 bars to variations in ammonia. We can also test the other end-case member by assuming that there is no ammonia variation below 10 bar as predicted by the simulations, and instead retrieve a physical temperature anomaly that matches the observed brightness temperature anomaly. 

\subsection*{Non-hydrostatic Simulations}

We utilize the non-hydrostatic cloud-resolving model to study Jupiter's atmosphere's large-scale behavior. We conduct simulations on a beta plane using altitude coordinates in a Cartesian box to simulate the differential rotation at different latitudes on Jupiter. The domain is bounded by solid walls in latitude at $\rm 20^\circ$ as the southern boundary and about $\rm 57^\circ$ as the northern boundary. We adopt doubly-periodic boundary conditions in longitude. The size of the domain is $4.5 \times 10^4$ km in latitude, $6\times 10^4$ km in longitude, and 340 km in altitude, in which the corresponding pressure range is about 2.5E-3 to 90 bar. We initialize the simulation with three times solar water and ammonia and initialize the numerical experiment with the corresponding thermochemical equilibrium moist adiabat \cite{Li2018}. The initial horizontal wind profiles are uniformly zero, but we implement a random and small vertical wind field, which is on the order of 1 $\rm cm\;s^{-1}$ to break the initial asymmetry. Convection is driven by the spatially uniform and temporally constant cooling at the top and heating at the bottom, and their fluxes are fixed to 7.5 $\rm W\;m^{-2}$, which is the recently corrected internal heat flux of Jupiter \cite{LiLim2018}. The heating and cooling scheme was adopted for many numerical simulations to study the general behavior of moist convection on giant planets \cite{Sugiyama2011,Sugiyama2014}. We slightly overforce the model by amplifying the heat flux by a factor of four in the first 3,000 days and then reducing the heat flux to the normal Jupiter value so the simulation can reach the statistically steady state faster.

The simulation produces several self-organized mid-latitude jets, belts, and zones. It also produces storms that periodically occur every three to four years. In this study, we chose the storm that happened around 4,000 simulation days to best match the ammonia and temperature retrieval from the Juno data.
% If your text is very short you might need to uncomment the following line to avoid
% layout problems with the figures and tables.
%\newpage

%%%%%%%%%%%%%%%% REFERENCES %%%%%%%%%%%%%%%

\clearpage % Clear all remaining figures and tables then start a new page

% The list of references goes after the main text and before the acknowledgements
% When preparing an initial submission, we recommend you use BibTeX, like this:
%

\bibliography{science_final} % for a file named science_template.bib
\bibliographystyle{sciencemag}

% After the paper has completed peer review and been revised ready for acceptance,
% you should comment out the lines above and copy-paste the contents of your .bbl
% file here instead. This will help ensure that our conversion software works correctly.
% Remember to re-run BibTeX first - check the timestamp!
%
% Example of the first three entries copy-pasted from science_template.bbl:
%
%\begin{thebibliography}{1}
%
%\bibitem{example}
%A.~N. {Author}, An example reference. \emph{Journal of Improbable Research}
%  \textbf{1}, 67 (2020).
%
%\bibitem{example2}
%F.~M. {Surname}, S.~{Author}, A second example. \emph{Interesting Research
%  Letters} \textbf{32}, 897 (2019).
%
%\bibitem{example_preprint}
%P.~{One}, P.~{Two}, P.~{Three}, {An unpublished preprint}. \emph{preprint}
%  (2021), arXiv:2101.12345.
%
%\end{thebibliography}

%%%%%%%%%%%%%%%% ACKNOWLEDGEMENTS %%%%%%%%%%%%%%%

\section*{Acknowledgments}
We gratefully acknowledge Adil Hussain for his early exploration of this storm system during his undergraduate research, where he studied the cloud structure using the correlated-k method. While the method proved unsuitable for this specific application, his efforts were instrumental in guiding the development of the deconvolution algorithm, which became a cornerstone of this work.

\paragraph*{Funding} 
Chris Moeckel and Imke de Pater were in part supported by the NASA’s Solar System Observations (SSO) award 80NSSC18K1003

\paragraph*{Author contributions:\newline } 
	Conceptualization: Chris Moeckel, Imke de Pater \\ 
	Methodology: Chris Moeckel, Imke de Pater\\ 
	Investigation: Chris Moeckel, Huazhi Ge \\  
	Visualization: Chris Moeckel \\  
	Supervision: Imke de Pater \\ 
	Data curation: Chris Moeckel, Huazhi Ge, Imke de Pater \\ 
	Formal analysis: Chris Moeckel, Huazhi Ge \\ 
	Software: Chris Moeckel, Huazhi Ge \\ 
	Writing—review and editing: Chris Moeckel, Imke de Pater \\  
	Writing—original draft: Chris Moeckel \\ 
	Resources: Imke de Pater \\ 
	Funding acquisition: Imke de Pater \\ 
	Project administration: Chris Moeckel, Imke de Pater\\ 
    Validation: Chris Moeckel, Imke de Pater \\

\paragraph*{Competing interests} 
``There are no competing interests to declare.''

\paragraph*{Data and materials availability:}
All data needed to evaluate the conclusions in the paper are present in the paper and/or the Supplementary Materials. The raw data of the Juno MWR instrument can be accessed under \url{https://pds-atmospheres.nmsu.edu/cgi-bin/getdir.pl?volume=jnomwr_1100}. The fully open and independent data pipeline to reduce the data and produce the deprojected maps can be accessed under \url{https://github.com/cmoeckel91/pyPR/blob/master/JunoTools.py}, and the reduced data are also available under Zenodo \url{10.5281/zenodo.14908671}. 

%%%%%%%%%%%%%%%% SUPPLEMENT LIST %%%%%%%%%%%%%%%

% List the contents of your Supplementary Materials, including the numbers of any
% supplementary figures, tables, external data files etc. and any references that are
% cited only in the supplement. In this example, refs. 7-8 are cited only in the supplement.
% Fill out your numbers accordingly and delete any lines that aren't applicable.
\subsection*{Supplementary materials}
Figs. S1 to S5\\

% (filling out the other numbers automatically is possible but fiddly and liable to break)

%%%%%%%%%%%%%%%% END OF MAIN TEXT %%%%%%%%%%%%%%%

\newpage

%%%%%%%%%%%%%%%% MAIN TEXT FIGURES %%%%%%%%%%%%%%%
\clearpage  

\clearpage  
    
\clearpage  
    
\clearpage 
\clearpage 
%%%%%%%%%%%%%%%% MAIN TEXT TABLES %%%%%%%%%%%%%%%

%%%%%%%%%%%%%%%% START OF SUPPLEMENT %%%%%%%%%%%%%%%

% Figures, tables, equations and pages in the supplement are numbered S1, S2 etc.
\renewcommand{\thefigure}{S\arabic{figure}}
\renewcommand{\thetable}{S\arabic{table}}
\renewcommand{\theequation}{S\arabic{equation}}
\renewcommand{\thepage}{S\arabic{page}}
\setcounter{figure}{0}
\setcounter{table}{0}
\setcounter{equation}{0}
\setcounter{page}{1} % not 0 as \newpage already started a supplementary page
% References continue the numbering from the main text.

%%%%%%%%%%%%%%%% SUPPLEMENT TITLE PAGE %%%%%%%%%%%%%%%

\begin{center}
\section*{Supplementary Materials for\\ \scititle}

% Author list for the supplement
% Indicate the corresponding authors, but do NOT include institutions here
% It would be nice if the template auto-generated this, but doing so is complicated...
Chris~Moeckel$^{\ast\dagger}$,
H.~Ge$^\dagger$,
I.~de Pater\\ % we're not in a \author{} environment this time, so use \\ for a new line
\small$^\ast$Corresponding author. Email: chris.moeckel@berkeley.edu\\
\small$^\dagger$These authors contributed equally to this work.
\end{center}

% Fill out the numbers for each type of supplementary material,
% and delete any lines that aren't applicable.
% These are just example numbers that don't match the rest of this template.
\subsubsection*{This PDF file includes:}
Materials and Methods\\
Supplementary Text\\
Figures S1 to S5\\

\newpage

%%%%%%%%%%%%%%%% MATERIALS AND METHODS %%%%%%%%%%%%%%%

%%%%%%%%%%%%%%%% SUPPLEMENTARY TEXT %%%%%%%%%%%%%%%

% If your supplement is very short you might need to uncomment the following line to avoid
% layout problems with the figures and tables.
%\newpage

%%%%%%%%%%%%%%%% SUPPLEMENTARY FIGURES %%%%%%%%%%%%%%%

\begin{figure}
\centering
\includegraphics[width=1\textwidth]{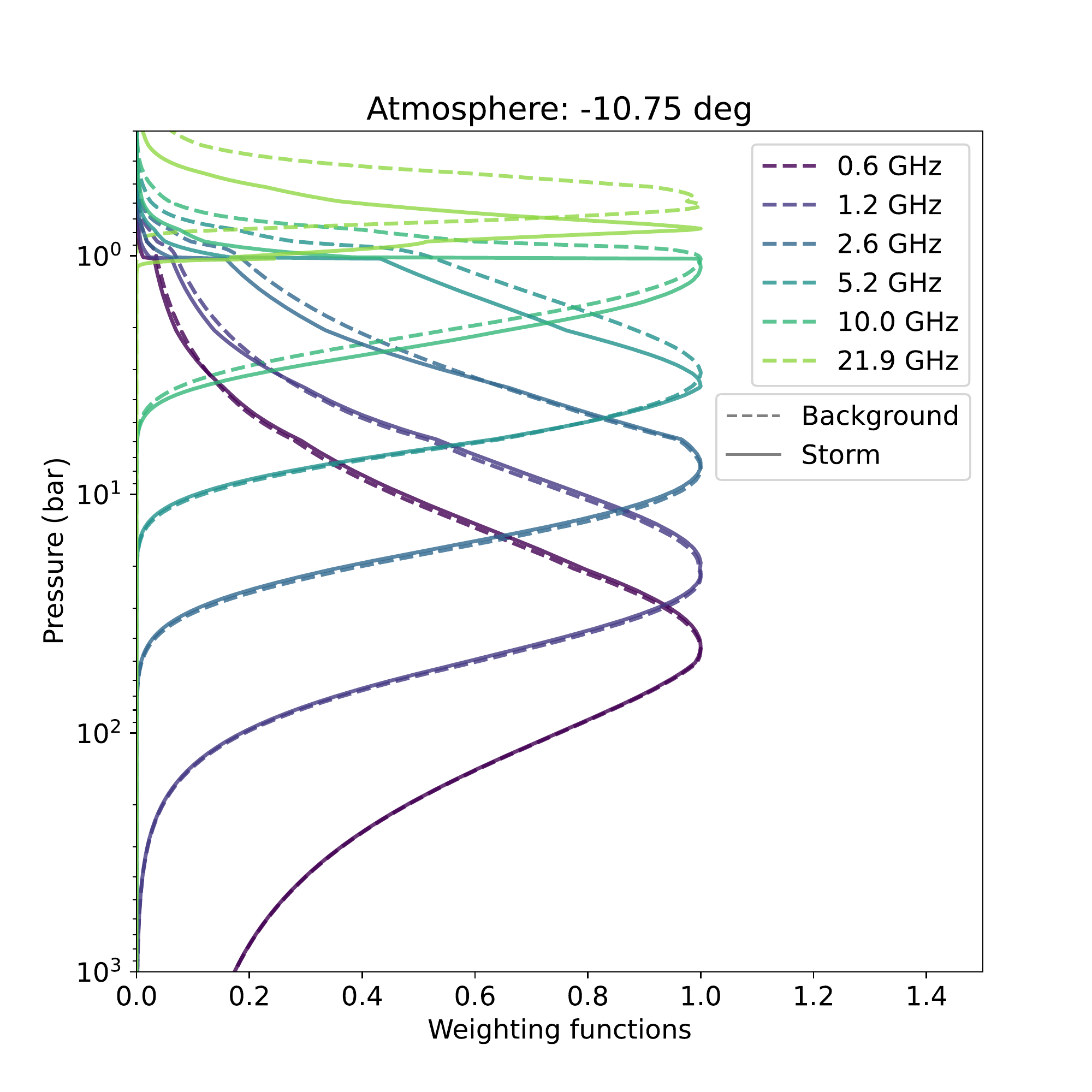}
\caption{\textbf{Weighting functions that represent the depths the emission originates from.} Radio signals record the opacity-weighted thermal emission by the atmosphere. Since the ammonia is the main factor controlling the opacity of the atmosphere, changes in the ammonia abundance will change the pressures from which the thermal signal is received. The dashed lines indicate the pressure from which the thermal emission is received for each frequency based upon the background ammonia profiles. The solid lines represents a storm profile (corresponding to the solid red line in Fig. \ref{fig:RelRetrieval}) and the dashed line corresponds to ammonia cloud region (solid blue line in Fig. \ref{fig:RelRetrieval}). The depleted upper atmosphere allows us to probe slightly deeper than compared to the background atmosphere.} 
\label{fig:WF}
\end{figure}

\clearpage  

\begin{figure}
\centering
\includegraphics[width=1\textwidth]{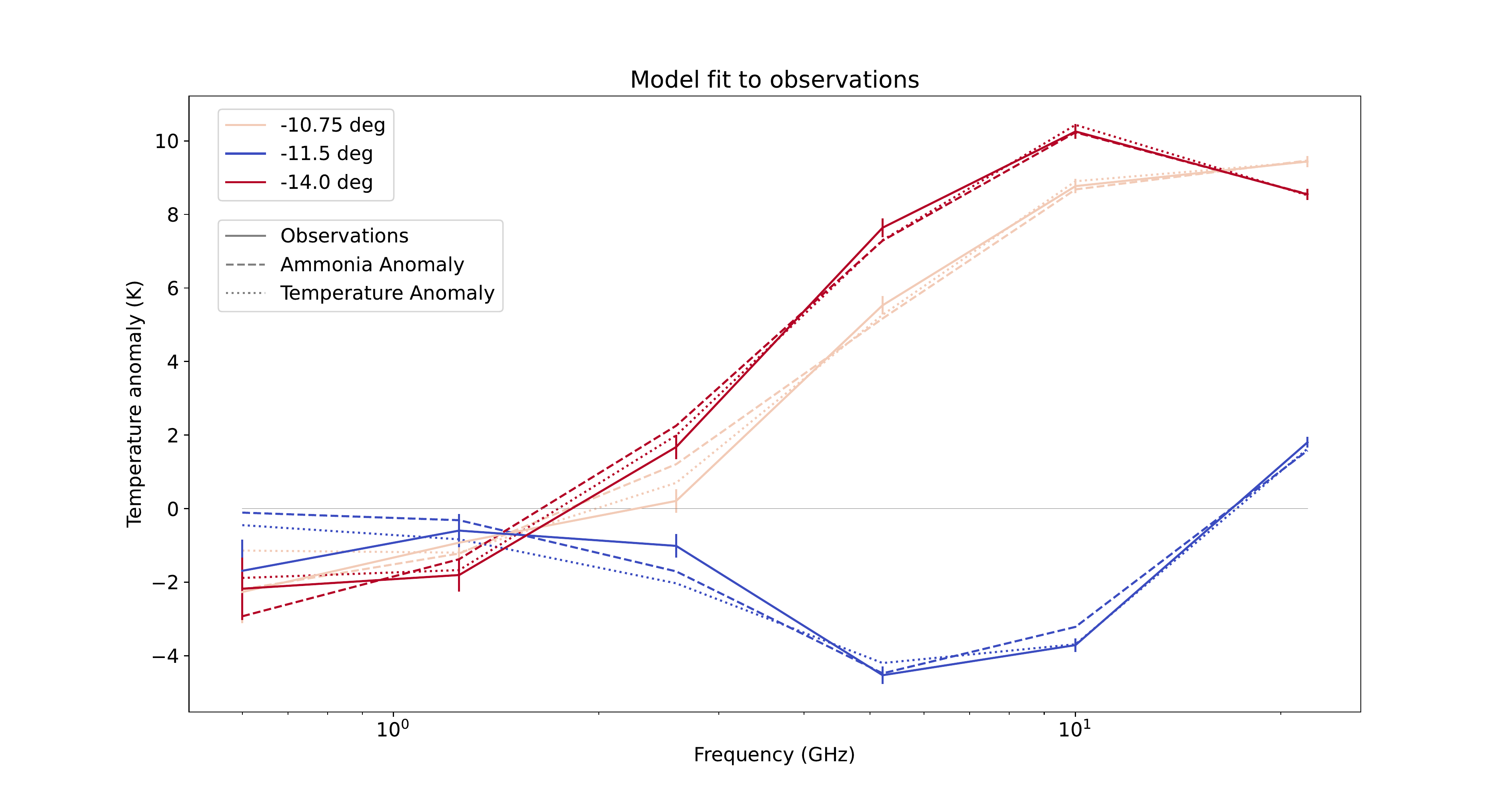}
\caption{\textbf{Comparison of observed and modeled brightness temperature anomaly}. The solid lines represents the observed brightness temperature anomaly for the three regions under consideration in Fig. \ref{fig:RelRetrieval}. The dashed line is the modeled brightness temperature difference between the background atmosphere and the storm atmosphere assuming a pure ammonia anomaly. The dotted line shows the result for keeping the ammonia abundance constant and instead showing a pure temperature anomaly. Both models, despite very different assumptions, fit the observations well.} 
\label{fig:AnomalyFit}
\end{figure}

\clearpage  

\begin{figure}
\centering
\includegraphics[width=\textwidth]{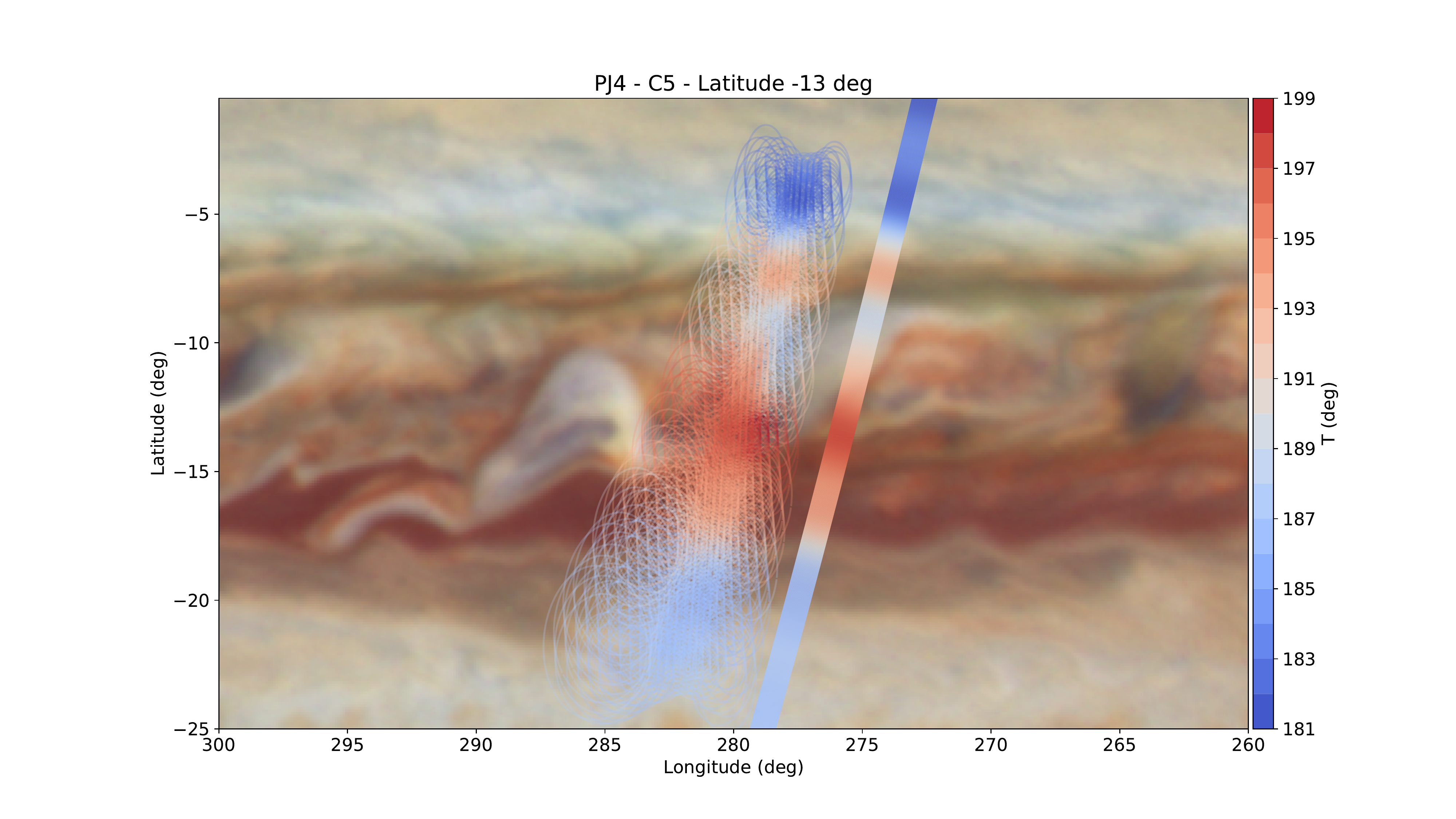}
\caption{\textbf{Example of the distribution of the available individual observations.} Each individual ellipse represents a single integration for the PJ4 flyby in the southern hemisphere. For displaying purposes alone, we have converted each measurement into nadir brightness temperature, by assuming a limb-darkening coefficient. The bar on the right hand side indicates the zonally averaged brightness temperature, and the starting point for the deconvolution algorithm.}
\label{fig:C5Footprints}
\end{figure}

\clearpage  

\begin{figure}
\centering
\includegraphics[width=0.5\textwidth]{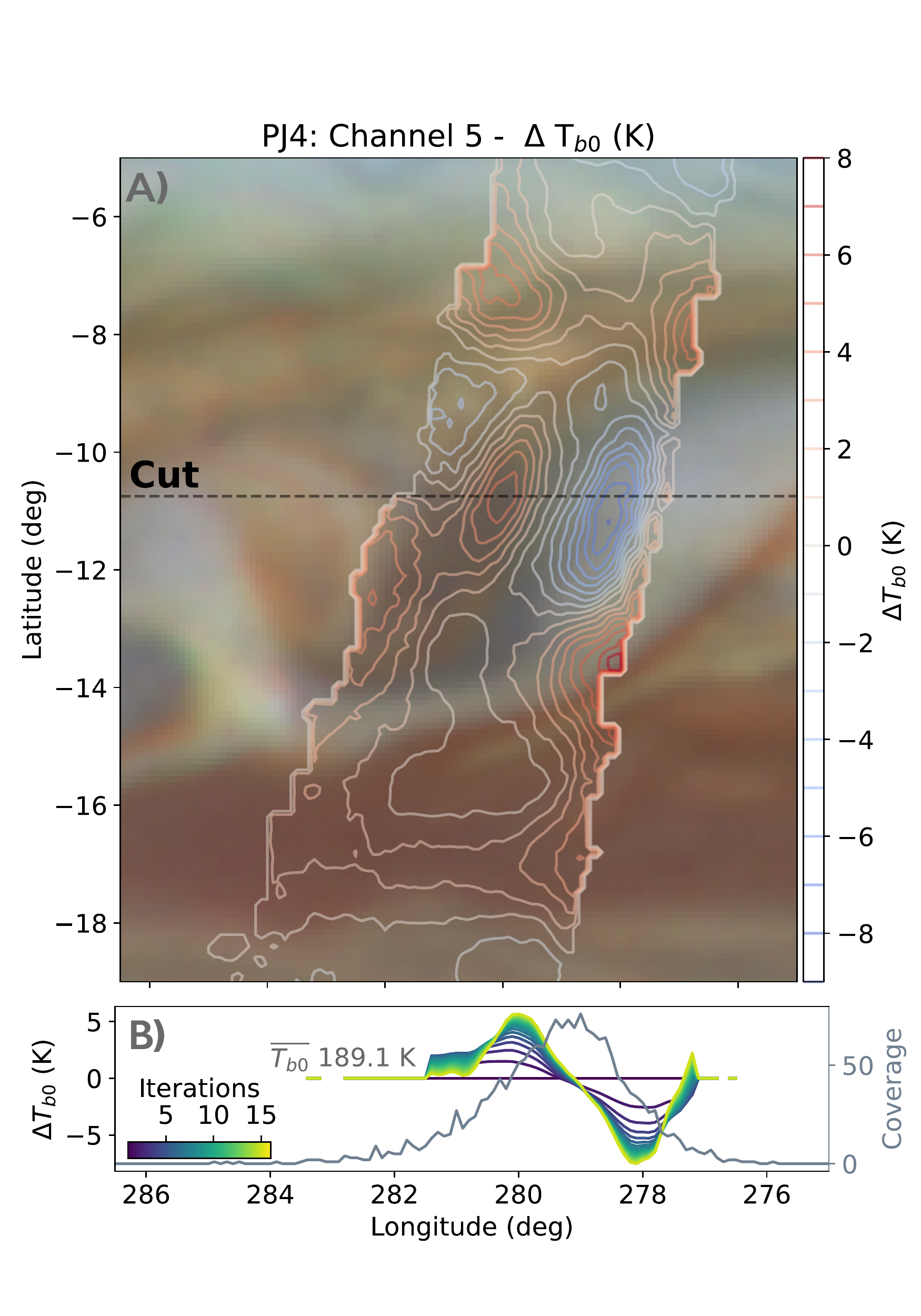}
\caption{\textbf{Example residual distribution for C5 during the PJ4 flyby}. Panel A) shows shows the $\Delta T_{b0}$ distribution overlain on concurrent HST observations. The retrieved residuals are largest in regions where the HST observations show a localized white storm plumes next to a darker background. Panel B) is a cross cut through the  $\Delta T_{b0}$ map where each line corresponds to the summed brightness residual after the number of indicated iterations, and the secondary axis indicates the number of observations for a given region of the map. }
\label{fig:PJ4-DTMap}
\end{figure}

\clearpage  

\begin{figure}
\centering
\includegraphics[width=0.6\textwidth]{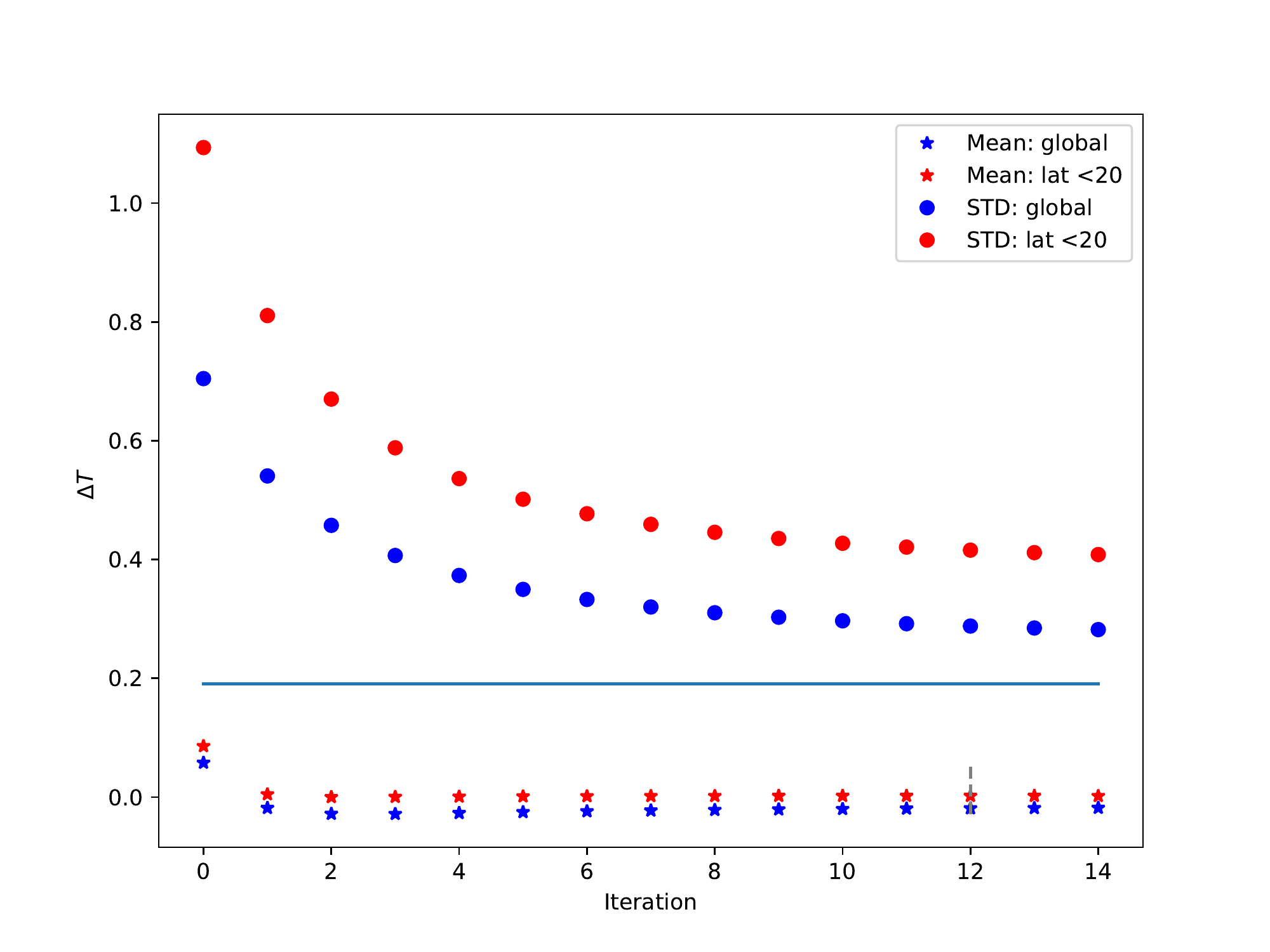}
\caption{\textbf{Convergence of the deconvolution algorithm}. The lines represent the mismatch for PJ4 - C5 across all measurements with the theoretical measurement error approach 0.1\% \cite{janssen2017} indicated by the horizontal line. We can see that within the first few iterations the standard deviations drops quickly until it slowly converges towards the theoretical limit. We track convergence in the tropical regions (lat $<$ 20 deg) and globally, to look for possible contamination by synchrotron radiation. }
\label{fig:Convergence}
\end{figure}
\clearpage  

%%%%%%%%%%%%%%%% SUPPLEMENTARY TABLES %%%%%%%%%%%%%%%

% \begin{table} % Do not use \begin{table*}
% 	\centering
% 	% Captions go above tables
% 	\caption{\textbf{All captions must start with a short bold sentence, acting as a title.}
% 		Follow the same style as main text tables.
% 		If the design is similar to previous tables, avoid repetition by refering back to them.}
% 	\label{tab:sup_example} % give each table a logical label name

% 	\begin{tabular}{lccr} % four columns, alignment for each
% 		\\
% 		\hline
% 		A & B & C & D\\
% 		\hline
% 		1 & 2 & 3 & 4\\
% 		2 & 4 & 6 & 8\\
% 		3 & 5 & 7 & 9\\
% 		\hline
% 	\end{tabular}
% \end{table}

%%%%%%%%%%% CAPTIONS FOR OTHER SUPPLEMENTARY FILES %%%%%%%%%%

\clearpage % Clear all remaining figures and tables then start a new page

% \paragraph{Caption for Movie S1.}
% \textbf{All captions must start with a short bold sentence, acting as a title.}
% Then explain what is shown in the supplementary video file.
% Give as much detail as you would for a figure e.g. explain axes, color maps etc.
% If the video is an animated equivalent of one of the static figures, state e.g.
% `Animated version of Figure~\ref{fig:example}.'

% \paragraph{Caption for Data S1.}
% \textbf{All captions must start with a short bold sentence, acting as a title.}
% Then explain what is included in the supplementary data file.
% Give as much detail as you would for a table e.g. explain the meaning of every column,
% units used, any special notation etc.

%%%%%%%%%%%%%%%% SUPPLEMENTARY REFERENCES %%%%%%%%%%%%%%%

% Do NOT include a reference list in the supplement.
% All references must be in a single list at the end of the main text.
% The copyeditors will ensure that the correct reference list appears with each version of the paper
% (print, HTML, PDF, mobile app, metadata for bibliographic databases etc.)

\end{document}